\documentclass{article}
\usepackage[english]{babel}
\usepackage{mydefs}
\usepackage{todonotes}

\title{A General Form of Attribute Exploration}
\author{
  \small Daniel Borchmann\\
  \small TU Dresden\\
  \small Faculty of Mathematics and Sciences\\
  \small Institute for Algebra\\
  \small \texttt{daniel.borchmann@mailbox.tu-dresden.de}}
\date{\today}

\begin{document}

\maketitle

\begin{abstract}
  We present a general form of attribute exploration, a knowledge completion algorithm
  from Formal Concept Analysis.  The aim of our presentation is not only to extend the
  applicability of attribute exploration by a general description.  It may also allow to
  view different existing variants of attribute exploration as instances of a general
  form, which may simplify theoretical considerations.
\end{abstract}


\section{Introduction}

Attribute exploration is a well known algorithm within formal concept
analysis~\cite{fca-book}.  Its main application can be summarized as \emph{semi-automatic
  knowledge base completion}.  Within this process, a domain expert is asked about the
validity of certain implications in the domain of discourse.  Based upon the answer of the
domain expert, the algorithm enhances its knowledge until all implications are known to
hold or not to hold in the domain, and the algorithm stops.

Attribute exploration has gained much attention since its first formulation, and for
certain problems variations of attribute exploration have been devised where the original
algorithm was not applicable.  Those variations include attribute exploration on partial
context~\cite{BGSS07} and exploration of models of the description logic
$\mathcal{EL}$~\cite{BaaderDistel08,BaDi09}, among others.

However, in almost all variations of attribute exploration that have been devised the
overall structure of the algorithm remains the same.  Furthermore, all important
properties of attribute exploration remain, and one might be tempted to ask whether a
general form of attribute exploration can be found that subsumes all many of these
variations.  The purpose of this work is to present some first considerations into this
direction.

We shall proceed as follows.  After introducing the mandatory definitions in the first
section we briefly revisit the classical description of attribute exploration as it is
given in~\cite{fca-book}.  Starting from this, we motivate our generalizations and
summarize the resulting algorithm together with its properties in the succeeding section.
We shall have a close look at a special cases which involves \emph{pseudoclosed sets} and
results in some very nice results about the attribute exploration algorithm.  Finally, we
shall summarize our considerations and give an outlook on further questions.

\section{Preliminaries}

As attribute exploration is an algorithm from Formal Concept Analysis, we shall begin by
introducing some basic definitions from within this field.  This includes notions like
formal contexts, contextual derivations, implications, partial contexts and pseudoclosed
sets.  We shall furthermore recall the notion of closure operators on sets, which we need
for our considerations.

Let $G$ and $M$ be two sets and let $I \subseteq G \times M$.  Then the triple $\con{K} :=
(G, M, I)$ is called a \emph{formal context}.  We shall connect with it the following
interpretation: The set $G$ is the set of \emph{objects} of $\con{K}$, $M$ is the set of
\emph{attributes} of $\con{K}$ and $(g,m)$ is an element of the incidence relation $I$ if
and only if the object $g$ \emph{has} the attribute $m$.  We may also write $g\rel{I}m$ if
$(g,m) \in I$. If $\con{K}$ is a formal context, then the set of objects, attributes and
the incidence relation is denoted by $G_\con{K}$, $M_\con{K}$ and $I_\con{K}$,
respectively.

Let us fix a formal context $\con{K} = (G, M, I)$.  If $A \subseteq G$, then the set of
\emph{common attributes of $A$ in $\con{K}$} is denoted by
\begin{align*}
  A' &:= \set{m \in M \mid \forall g \in A: g\rel{I}m}
\intertext{and likewise for $B \subseteq M$,}
  B' &:= \set{g \in G \mid \forall m \in B: g\rel{I}m}
\end{align*}
denotes the set of all \emph{common objects of $B$ in $\con{K}$}.  The sets $A'$ and $B'$
are called the \emph{(contextual) derivations} of the respective sets, and the operators
named $(\cdot)'$ are hence called the \emph{derivation operators} of $\con{K}$.
\begin{Lemma}
  Let $\con{K} = (G, M, I)$ be a formal context and let $A, A_1, A_2 \subseteq M, B, B_1,
  B_2 \subseteq G$.  Then the following statements hold:
  \begin{enumerate}[i) ]
    \item $A_1 \subseteq A_2 \implies A_2' \subseteq A_1'$
    \item $B_1 \subseteq B_2 \implies B_2' \subseteq B_1'$ 
    \item $A \subseteq A''$
    \item $B \subseteq B''$
    \item $A' = A'''$
    \item $B' = B'''$
    \item $A \subseteq B' \iff A' \supseteq B$
  \end{enumerate}
\end{Lemma}

As we view the elements of $G$ as objects with certain attributes from $M$, we may ask for
two sets $A, B \subseteq M$ whether all objects having all attributes from $A$ also have
all attributes from $B$.  This can be rewritten in terms of the derivations operators as
$A' \subseteq B'$.  We shall call the pair $(A, B)$ an \emph{implication on $M$} and
denote it as $A \to B$.  If $\con{K}$ is a formal context with attribute set $M$, then we
may also say that $A \to B$ is an \emph{implication of $\con{K}$}.  Then $A$ is called the
\emph{premise} and $B$ the \emph{conclusion} of the implication.  If indeed $A' \subseteq
B'$, we shall call $A \to B$ a \emph{valid} implication of $\con{K}$, and we may write
$\con{K} \models (A \to B)$.  As $A' \subseteq B' \iff B \subseteq A''$, we can observe
that
\begin{equation*}
  \con{K} \models (A \to B) \iff B \subseteq A''.
\end{equation*}
We shall denote with $\Imp(M)$ the set of all implications on $M$, with $\Imp(\con{K})$
the set of all implications of $\con{K}$ and with $\Th(\con{K})$ the set of all valid
implications of $\con{K}$.

Let $\mathcal{L} \subseteq \Imp(\con{K})$ and let $A \subseteq M$.  The set $A$ is
\emph{closed under $\mathcal{L}$} if for all implications $(X \to Y) \in \mathcal{L}$ it
holds that $X \not\subseteq A$ or $Y \subseteq A$.  Let us further define
\begin{align*}
  \mathcal{L}^0(A) &:= A,\\
  \mathcal{L}^1(A) &:= \bigcup\set{Y \mid (X \to Y) \in \mathcal{L}, X \subseteq A},\\
  \mathcal{L}^i(A) &:= \mathcal{L}^1(\mathcal{L}^{i-1}(A))\quad\text{for $i > 1$},
\intertext{and}
  \mathcal{L}(A) &:= \bigcup_{i \in \NN}\mathcal{L}^i(A).
\end{align*}
The set $\mathcal{L}(A)$ is then the smallest superset of $A$ that is closed under
$\mathcal{L}$.

The set $\Th(\con{K})$ might be quite large, and to handle this set in practical
applications it is desirable to represent it by a small subsets.  To see how this is done
let $\mathcal{L} \subseteq \Imp(\con{K})$ and let $(A \to B) \in \Imp(\con{K})$.  Then
$\mathcal{L}$ \emph{entails} $A \to B$, written as $\mathcal{L} \models (A \to B)$, if and
only if $B \subseteq \mathcal{L}(A)$.  A set $\mathcal{B} \subseteq \Imp(\con{K})$ is
called \emph{sound} for $\mathcal{L}$ if every implication from $\mathcal{B}$ is entailed
by $\mathcal{L}$.  $\mathcal{B}$ is said to be \emph{complete} for $\mathcal{L}$ if every
implication from $\mathcal{L}$ is entailed by $\mathcal{B}$.  If $\mathcal{B}$ is both
sound and complete for $\mathcal{L}$, it is called a \emph{base} for $\mathcal{L}$.  It is
called a \emph{non-redundant base} for $\mathcal{L}$ if it is $\subseteq$-minimal with
respect to this property.

Let us denote with $\Cn(\mathcal{L})$ the set of all implications that are entailed by
$\mathcal{L}$.  Then
\begin{align*}
  \text{$\mathcal{B}$ is sound for $\mathcal{L}$} &\iff \mathcal{B} \subseteq
  \Cn(\mathcal{L}),\\
  \text{$\mathcal{B}$ is complete for $\mathcal{L}$} &\iff \Cn(\mathcal{B}) \supseteq
  \mathcal{L}.
\end{align*}
In particular, $\mathcal{B}$ is a base for $\mathcal{L}$ if and only if $\Cn(\mathcal{B})
= \Cn(\mathcal{L})$.

From all possible bases for $\mathcal{L}$ one can explicitly describe a \emph{canonical
  base} for $\mathcal{L}$ which has the remarkable property that it has minimal
cardinality among all bases for $\mathcal{L}$.  Let $P \subseteq M$.  Then $P$ is said to
be \emph{pseudoclosed under $\mathcal{L}$} if
\begin{enumerate}
\item $P \neq \mathcal{L}(P)$ and
\item for all pseudoclosed sets $Q \subsetneq P$ it follows $\mathcal{L}(Q) \subseteq P$.
\end{enumerate}
In particular, if $\mathcal{L} = \Th(\con{K})$, then $P$ is said to be a
\emph{pseudointent of $\con{K}$}.  Now the \emph{canonical base} for $\mathcal{L}$ is
defined as
\begin{equation*}
  \Can(\mathcal{L}) := \set{P \to \mathcal{L}(P) \mid P \text{ pseudoclosed under $\mathcal{L}$}}.
\end{equation*}

Formal contexts require a certain kind of complete knowledge about their objects:  If
$g \in G$ and $m \in M$ then either $g$ has the attribute $m$ or not.  Under certain
circumstances this might be inappropriate, because it might not be known whether $g$ has
the attribute $m$, or it is simply irrelevant for the task at hand.  Therefore we shall
introduce \emph{partial contexts}.

Let $M$ be a set.  Then a partial context $\con{K}$ is a set of pairs $(A,B)$ with
$A,B \subseteq M$ such that $A \cap B = \emptyset$.  Such a pair is called a \emph{partial
  object description} if $A \cup B \neq M$ and a \emph{full object description} if $A \cup B
= M$.  Intuitively, one can understand partial objects descriptions as a pair of positive
attributes, \ie attributes the corresponding object definitively has, and negative
attributes, \ie attributes the corresponding object definitively does not have.  The
objects itself are not named in partial contexts.

An \emph{implication} for $\con{K}$ is just an implication on $M$.  Such an implication
$(A \to B) \in \Imp(M)$ is refuted by $\con{K}$ if there exists a partial object
description $(X,Y) \in \con{K}$ such that $A \subseteq X, B \cap Y \neq \emptyset$.  If $A
\subseteq M$, then the $\subseteq$-maximal set $B$ such that $A \to B$ is not refuted by
$\con{K}$ exists and is given by
\begin{equation*}
  \con{K}(A) := B := M \setminus \bigcup\set{Y \mid (X,Y) \in \con{K}, A \subseteq X}.
\end{equation*}

As it turns out, the operators $(\cdot)''$, $\mathcal{L}(\cdot)$ and $\con{K}(\cdot)$ are
instances of the more abstract notion of \emph{closure operators on sets}.  Let again $M$
be a set.  Then a function $c \colon \subsets{M} \to \subsets{M}$ is said to be a
\emph{closure operator on $M$} if and only if
\begin{enumerate}[i) ]
\item $c$ is \emph{extensive}, \ie $A \subseteq c(A)$ for all $A \subseteq M$,
\item $c$ is \emph{idempotent}, \ie $c(c(A)) = c(A)$ for all $A \subseteq M$,
\item $c$ is \emph{monotone}, \ie $A \subseteq B \implies c(A) \subseteq c(B)$ for all $A,
  B \subseteq M$.
\end{enumerate}
Both $(\cdot)''$ and $\mathcal{L}(\cdot)$ are closure operators on their corresponding sets
of attributes.  A set $A \subseteq M$ is said to be \emph{closed under $c$} if $c(A) = A$.
The set of all closed sets of $c$, \ie the image of $c$, is denoted by $\im c$.  A set
$P \subseteq M$ is said to be \emph{pseudoclosed under $c$} if and only if
\begin{enumerate}[i) ]
\item $P \neq c(P)$ and
\item for all pseudoclosed $Q \subseteq P$, it holds that $c(Q) \subseteq P$.
\end{enumerate}

We shall write $c_1(\cdot) \subseteq c_2(\cdot)$ for two closure operators $c_1, c_2$ on a
set $M$ if and only if $c_1(A) \subseteq c_2(A)$ for all $A \subseteq M$.

\section{Classical Attribute Exploration}

Given a finite set $M$, attribute exploration semi-automatically tries to determine the
set of implications that are valid in a certain domain.  Together with a set $\mathcal{K}$
of already known valid implications and a formal context $\con{K}$ of valid examples,
attribute exploration generates implications $A \to B$ that hold in $\con{K}$ but are not
entailed by $\mathcal{K}$.  Those implications are asked to the expert for validity.  If
$A \to B$ holds in the domain of discourse, it is added to the set $\mathcal{K}$.
Otherwise the expert has to present a counterexample for $A \to B$ that is added to the
formal context $\con{K}$.  The procedure terminates if there are no such implications
left.

To describe attribute exploration more formally, let us define what is meant by a domain
expert.
\begin{Definition}
  Let $M$ be a set.  A \emph{domain expert on $M$} is a function
  \begin{equation*}
    p \colon \Imp(M) \to \set{\top} \cup \subsets{M},
  \end{equation*}
  where $\top$ is a special symbol not equal to any subset of $M$, such that the following
  conditions hold
  \begin{enumerate}[i) ]
  \item If $X \to Y$ is an implication on $M$ such that $p(X \to Y) = C \neq \top$, then
    $X \subseteq C, Y \not\subseteq C$. (\emph{$p$ gives counterexamples for false
      implications})
  \item If $A \to B$ and $X \to Y$ are implications on $M$ such that $p(A \to B) = \top$ and
    $p(X \to Y) = C \neq \top$, then $C$ is closed under $\set{A \to B}$, \ie
    $A \not\subseteq C$ or $B \subseteq C$. (\emph{counterexamples do not invalidate correct
      implications})
  \end{enumerate}
  If $p(A \to B) = \top$, then we say that \emph{$p$ confirms $A\to B$}.  Otherwise we
  say that $p$ \emph{rejects} the implication and we call the set $C = p(A \to B) \neq
  \top$ a \emph{counterexample from $p$ for $A\to B$}.  Finally, the \emph{theory of $p$}
  is just the set of implications that $p$ confirms, \ie
  \begin{equation*}
    \Th(p) := p^{-1}(\set{\top}) = \set{A \to B \mid p(A \to B) = \top}.
  \end{equation*}
\end{Definition}

An immediate consequence of the definition is the following observation.
\begin{Lemma}
  Let $\mathcal{L}$ be a set of implications such that a given domain expert $p$ confirms
  every implication in $\mathcal{L}$.  If $\mathcal{L} \models (A \to B)$, then $p$
  confirms $A \to B$ as well.
  \begin{Proof}
    Suppose that $p(A \to B) = C \neq \top$.  Then $C$ is closed under $\mathcal{L}$.  This
    means that $\mathcal{L}(C) = C$.  Since $\mathcal{L} \models (A \to B)$, from
    $A \subseteq C$ it follows that
    \begin{equation*}
      B \subseteq \mathcal{L}(A) \subseteq \mathcal{L}(C) = C,
    \end{equation*}
    \ie $C$ is not a counterexample for $A \to B$, a contradiction.
  \end{Proof}
\end{Lemma}

Before we are able to describe the attribute exploration algorithm more formally, we need
to give another definition.
\begin{Definition}
  Let $M$ be a finite set and let $<$ be a total order on $M$.  Then for $A,B \subseteq M$
  and $i \in M$ we define
  \begin{equation*}
    A \prec_i B \;:\!\iff i = \min\nolimits_<(A \Delta B),
  \end{equation*}
  where $A \Delta B = (A \setminus B) \cup (B \setminus A)$ is the symmetric difference
  between $A$ and $B$.  If $A \prec_i B$, we say that \emph{$A$ is lectically smaller than
    $B$ at $i$}.  Furthermore, \emph{$A$ is lectically smaller than $B$}, written as
  $A \prec B$, if there exists $i \in M$ such that $A \prec_i B$.  Finally,
  \begin{equation*}
    A \preceq B \iff A = B \text{ or } A \prec B.
  \end{equation*}
\end{Definition}
It is easy to see that $\preceq$ constitutes a linear ordering on $\subsets{M}$.

We are now able to describe the process of attribute exploration in a more formal way.
\begin{Algorithm}[Classical Attribute Exploration]
  \label{algorithm:classical-attribute-exploration}
  Let $M$ be a finite set, $\con{K}$ be a formal context with attribute set $M$ and let
  $\con{K} \subseteq \Imp(M)$ and let $p$ be a domain expert on $M$.  Suppose that
  $\mathcal{K} \subseteq \Th(p) \subseteq \Th(\con{K})$.
  \begin{enumerate}[i) ]
  \item Initialize $P$ to the lectically first closed set of $\mathcal{K}(\cdot)$.
  \item \label{ae:loop start} If $P'' = P$, then go to~\ref{ae:next closed set}.
    Otherwise let $r := (P \to P'')$.
  \item If $p$ confirms $r$, then add $r$ to $\mathcal{K}$.
  \item If $p$ gives a counterexample $C$ for $r$, add a new object to $\con{K}$ which has
    exactly the attributes in $C$.
  \item \label{ae:next closed set} Let $Q$ be the lectically next closed set after $P$ of
    $\mathcal{K}$.  If there is none left, terminate.  Otherwise, set $P$ to $Q$ and go
    to~\ref{ae:loop start}.
  \end{enumerate}
  In any iteration, the current value of $\mathcal{K}$ is called the set of
  \emph{currently known implications} and the current value of $\con{K}$ is called the
  \emph{current working context}.
\end{Algorithm}

A first easy observation for this algorithm is the following: Suppose the expert $p$ is
called with an implication $A \to B$ during the run of the algorithm.  Let $\mathcal{K}$
be the currently known implications at this time, and let likewise $\con{K}$ denote the
current working context.  Then for each $m \in B$ both $\Th(p) \models (P \to \set{m})$
and $\Th(p) \not\models (P \to \set{m})$ is possible.  In other words, the question
whether $\Th(p) \models (P \to \set{m})$ is not influenced by the values of $\mathcal{K}$
and $\con{K}$ but solely depends on how the expert $p$ answers.  Hence all questions to
the expert can be seen as \emph{non-redundant}.

This property is very important especially in the presence of human experts which may not
only be expensive to answer but might also get impatient when getting asked implications
the algorithm could have inferred by itself.  Therefore, this property should of course
also hold for our generalized formulation of the attribute exploration, and it does, as we
shall see.

But before we do so, we shall mark down some of the major properties of this attribute
exploration algorithm.

\begin{Theorem}
  Let $M$ be a finite set, $<$ a total order on $M$, $\con{K}$ a formal context with
  attribute set $M$, $\mathcal{K}$ a set of implications on $M$ and let $p$ be a domain
  expert on $M$, such that $p$ confirms $\mathcal{K}$ and all implications confirmed by
  $p$ hold in $\con{K}$, \ie $\mathcal{K} \subseteq \Th(p) \subseteq \Th(\con{K})$.
  \begin{enumerate}[i) ]
  \item The attribute exploration algorithm terminates with $\con{K}$, $\mathcal{K}$ and
    $p$ as input.
  \item Let $\mathcal{K}'$ and $\con{K}'$ be the values corresponding to $\mathcal{K}$ and
    $\con{K}$ after the last iteration of the attribute exploration algorithm.  Then
    $\mathcal{K}'$ is a base for $\Th(\con{K}')$.
  \item $\Th(p) = \Th(\con{K}')$ and the corresponding closure operator coincides with
    $\mathcal{K}'(\cdot)$.
  \item The cardinality of $\mathcal{K}'\setminus\mathcal{K}$ is the smallest possible.
  \item The premises in $\mathcal{K}'\setminus\mathcal{K}$ are the
    \emph{$\mathcal{K}$-pseudoclosed of $Th(\con{K}')$}.  Thereby, a set $P \subseteq M$
    is said to be $\mathcal{K}$-pseudoclosed under $\mathcal{L}$ for $\mathcal{K},
    \mathcal{L} \subseteq \Imp(M)$, if and only if
    \begin{enumerate}[i) ]
    \item $P = \mathcal{K}(P)$,
    \item $P \neq \mathcal{L}(P)$,
    \item for each $\mathcal{K}$-pseudoclosed set $Q \subsetneq P$ of $\mathcal{L}$ it
      holds that $\mathcal{L}(Q) \subseteq P$.
    \end{enumerate}
  \end{enumerate}
\end{Theorem}
All but the last statement of the theorem are known
from~\cite{fca-book,stumme96attribute,DBLP:journals/tcs/Ganter99}.  The last statement has
been mentioned partially in~\cite{stumme96attribute} and has been proven completely
in~\cite{Diss-Felix}.

\section{Generalizing Attribute Exploration}

We shall now proceed by investigating the above description of attribute exploration for
possible generalizations.  While doing so, we shall not only generalize certain aspect of
the algorithm but also generalize those aspects intuitively.  The main aim of our
generalization is to describe attribute exploration in more abstract terms, to allow
applications of the algorithm beyond those of the classical algorithm.

Let $p$ be a domain expert on a set $M$.  We start with an informal introduction of our
generalizations, of which we shall name three:
\newcommand{\ccert}{c_\mathrm{cert}}
\newcommand{\cuniv}{c_\mathrm{univ}}
\begin{enumerate}[1. ]
\item The use of the initial formal context $\con{K}$ and the background knowledge
  $\mathcal{K}$ can be reduced to their corresponding closure operators $(\cdot)''$ and
  $\mathcal{K}(\cdot)$.  The only major problem here is the handling of counterexamples,
  which we shall discuss latter in detail.  Hence instead of passing the attribute
  exploration algorithm a formal context and some background knowledge in the form of a
  set of valid implications, we instead provide two closure operators $\cuniv$ and
  $\ccert$ on the set $M$.

  The closure operator $\cuniv$ takes the place of $\Th(\con{K})(\cdot)$ and represents
  the \emph{universal knowledge} we already have about our domain of discourse.  If
  $A \subseteq M$ is a set of attributes, then $\cuniv(A)$ represents the attributes that
  \emph{can} follow from $A$.  Seen from another perspective, $M\setminus\cuniv(A)$ is the
  set of attributes that \emph{do not follow} from $A$.

  In contrast to this, the closure operator $\ccert$ represents the \emph{certain}
  knowledge we already have.  In other words, $\ccert(A)$ is the set of all attributes
  that \emph{definitively follow} from $A$.  This closure operators hence takes the place
  of the set $\mathcal{K}$ of initially known implications.

  Clearly, we need to have $\ccert(\cdot) \subseteq \Th(p)(\cdot) \subseteq
  \cuniv(\cdot)$.
\item When providing counterexamples, we observe that we actually do not need to
  \emph{completely specify} them.  It merely is sufficient to provide information on which
  attributes a certain object has and which it not, as long as this information
  contradicts a proposed implication.  We shall take this approach and extend the
  algorithm to store those counterexamples in a partial context.  This idea has also been
  discussed in~\cite{BGSS07,GORS-book}.
\item The implications which are proposed to the expert are of a very special form, which
  guarantees certain optimality statements about the algorithm.  However, for the main
  application of knowledge acquisition and knowledge completion, this rather special form
  can be viewed as a certain kind of optimization.  To drop it, we may rather say that in
  any iteration step of the attribute exploration algorithm, we search for an
  \emph{undecided implication} with respect to the current values of $\ccert$ and
  $\cuniv$, \ie an implication $A \to B$ on $M$ such that $\ccert(A) \subsetneq B
  \subseteq \cuniv(A)$ and where both $A$ and $B$ are finite.  For such an implication we
  cannot infer from $\ccert$ and $\cuniv$ whether attributes $\cuniv(A) \setminus B$
  follow from $A$ or not, and hence we have to ask the domain expert.
\end{enumerate}
We shall take these observations as guidelines for our further considerations.  We start
by generalizing our notion of a domain expert such that we allow partial counter examples.
Next we present and discuss our general form of attribute exploration that incorporates
the above mentioned ideas.  For this we shall also prove correctness and non-redundancy of
the questions asked to the expert.  Subsequently, we shall have a closer look on how to
compute undecided implications in our general setting as it is done in the classical case.

\begin{Definition}
  Let $M$ be a set.  A function $q \colon \Imp(M) \to \set{\top} \cup \subsets{M}^2$ is
  said to be a \emph{partial domain expert on $M$} if and only
  if $\top$ is an element not in $\subsets{M}^2$ and the following conditions hold:
  \begin{enumerate}[1. ]
  \item If for $(A \to B) \in \Imp(M)$ it holds that $q(A \to B) = (C,D) \neq \top$, then
    $C \cap D = \emptyset$, $A \subseteq C$ and $B \cap D \neq \emptyset$. (\emph{$q$
      gives sufficient counterexamples for false implications})
  \item If $(A \to B), (X \to Y) \in \Imp(M)$ are such that $q(A \to B) = \top$ and $q(X
    \to Y) = (C, D) \neq \top$, then if $A \subseteq C$ then $B \cap D =
    \emptyset$. (\emph{counterexamples do not refute correct implications})
  \end{enumerate}
  As in the case for domain experts, we say that $q$ \emph{confirms} an implication $A \to
  B$ if and only if $q(A \to B) = \top$.  Otherwise we say that $q$ \emph{rejects} the
  implication and we call $q(A \to B) \neq \top$ a \emph{counterexample from $q$} for $A
  \to B$.  $\Th(q)$ shall denote the set of all implications on $M$ that are confirmed by
  $q$.
\end{Definition}
The counterexamples given by a partial domain expert can be seen as partial object
descriptions that are enough to invalidate a given implication.

Let us first investigate immediate consequences from the definition.  One of those is the
fact, as one would expect, that $\Th(q)$ is closed under entailment, \ie $\Cn(\Th(q)) =
\Th(q)$.

\begin{Lemma}
  \label{lemma:experts-are-consistent}
  Let $\mathcal{L} \subseteq \Imp(M)$ for a set $M$ and let $q$ be a partial domain expert
  on $M$, such that $q$ confirms all implications in $\mathcal{L}$.  If $\mathcal{L}
  \models (A \to B)$ for some $(A \to B) \in \Imp(M)$, then $q$ confirms $A \to B$ as
  well.
  \begin{Proof}
    Suppose that $q(A \to B) = (C, D)$ is a counterexample from $q$ for $A \to B$.  Then
    $A \subseteq C$.  Now $\mathcal{L}(C) \subseteq M \setminus D$ by the second condition
    on partial domain experts.  Since $\mathcal{L} \models (A \to B)$, it follows that $B
    \subseteq \mathcal{L}(A) \subseteq \mathcal{L}(C) \subseteq M \setminus D$.
    Therefore, $B \cap D = \emptyset$, contradicting the fact that $(C, D)$ is a
    counterexample for $A \to B$ from $q$.
  \end{Proof}
\end{Lemma}

\begin{Lemma}
  \label{lemma:counterexamples-are-closed-under-theory}
  If $(C, D)$ is a counterexample given by a partial domain expert $q$ on $M$, then
  $\Th(q)(C) \cap D = \emptyset$.
  \begin{Proof}
    By Lemma~\ref{lemma:experts-are-consistent}, $q$ confirms $C \to Th(q)(C)$.
    Therefore, by the second condition in the definition of $q$, it follows $D \cap
    \Th(q)(C) = \emptyset$, as required.
  \end{Proof}
\end{Lemma}

\begin{Lemma}
  \label{lemma:experts-and-partial-contexts}
  For a partial context $\con{K}$ with attribute set $M$ and a partial domain expert $q$
  on $M$ it holds that $\Th(q)(\cdot) \subseteq \con{K}(\cdot)$ if and only if $\Th(q)(C)
  \subseteq M \setminus D$ for each $(C, D) \in \con{K}$.
  \begin{Proof}
    $\Th(q)(\cdot) \subseteq \con{K}(\cdot)$ implies $\Th(q)(C) \cap D = \emptyset$ for
    each $(C, D) \in \con{K}$, which is equivalent to $\Th(q)(C) \subseteq M \setminus D$.

    For the converse let $\Th(q)(C) \cap D = \emptyset$ for all $(C, D) \in \con{K}$.  Let
    $A \subseteq M$.  Then for every $(C, D) \in \con{K}$ with $A \subseteq C$, it follows
    that $\Th(q)(A) \cap D \subseteq \Th(q)(C) \cap D = \emptyset$.  Therefore
    \begin{equation*}
      \Th(q)(A) \cap \bigcup\set{D \mid (C, D) \in \con{K}, A \subseteq C} = \emptyset
    \end{equation*}
    and hence $\Th(q)(A) \subseteq \con{K}(A)$ as required.
  \end{Proof}
\end{Lemma}

With those observations at hand, we are now able to state our generalized formulation of
the attribute exploration algorithm.
\begin{Algorithm}[General Attribute Exploration]
  \label{algorithm:general-attribute-exploration}
  Let $M$ be a set, $\ccert, \cuniv$ closure operators on $M$ and $q$ a partial domain
  expert $M$, such that $\ccert(\cdot) \subseteq \Th(q)(\cdot) \subseteq \cuniv(\cdot)$.
  \begin{enumerate}[i. ]
  \item Let $\con{K} = \emptyset$.
  \item \label{generalae:undecided-implication} Let $A \subseteq M$ be finite and such
    that there exists a finite set $B \subseteq M$ with $\ccert(A) \subsetneq B \subseteq
    \cuniv(A)$.  If there is no such set, terminate with output $\con{K}$ and $\ccert$.
    Otherwise consider the implication $A \to B$.
  \item If $q$ confirms $A \to B$, then update $\ccert$ to be the closure operators whose
    closed sets are exactly the closed sets of $\ccert$ that are also closed under $\set{A
      \to B}$.
  \item Otherwise let $(C, D) = q(A \to B)$ be a counterexample from $q$ for $A \to B$.
    Add $(C, D)$ to $\con{K}$.
  \item \label{generalae:update-counterexamples} Replace all counterexamples $(C, D) \in
    \con{K}$ by $(C', D')$, where
    \begin{align*}
      C' &:= \ccert(C), \\
      D' &:= D \cup \set{m \in M \setminus D \mid \ccert(C \cup \set{m}) \cap D \neq
        \emptyset}.
    \end{align*}
  \item Update $\cuniv$ to be the closure operator given by
    \begin{equation*}
      X \mapsto \cuniv(X) \cap \con{K}(X)
    \end{equation*}
    for all $X \subseteq M$.
  \item Go to~\ref{generalae:undecided-implication}.
  \end{enumerate}
\end{Algorithm}

Starting from this reformulation of the attribute exploration algorithm we shall now
consider the properties this algorithm has.  We shall show in this section that the
algorithm, as in the classical case, does not ask question its answers it could infer
itself.  Furthermore, the algorithm is correct in the sense that it returns a complete
description of the domain the given partial domain expert represents.  Termination,
however, cannot be shown in general, and we shall only give some sufficient condition.

The results in the minimality of the resulting set of confirmed implications does not hold
in this general setting.  For this, we have to generate the implications asked to the
expert in a way similar to the classical case.  We shall discuss this in more detail in
the next section.

To discuss the properties of Algorithm~\ref{algorithm:general-attribute-exploration}, we
need the following result.

\begin{Lemma}
  \label{lemma:ccert-subseteq-conK}
  At the end of every iteration of the generalized attribute exploration algorithm it
  holds that $\ccert(\cdot) \subseteq \Th(q)(\cdot) \subseteq \cuniv(\cdot)$ for the
  current values of $\ccert$ and $\cuniv$.  In particular, $\ccert(X) \subseteq
  \con{K}(X)$ holds for all $X \subseteq M$ at the end of every iteration.
  \begin{Proof}
    We prove the claim by induction.  For the base case we observe that $\con{K} =
    \emptyset$ and therefore $\con{K}(X) = M$ for all $X \subseteq M$.  Furthermore
    $\ccert(\cdot) \subseteq \Th(q)(\cdot) \subseteq \cuniv(\cdot)$ by the prerequisites
    of the algorithm.

    For the induction step assume that $\ccert(\cdot) \subseteq \Th(q)(\cdot) \subseteq
    \cuniv(\cdot)$ holds at the beginning of the current iteration.  Assume $A, B
    \subseteq M$ finite such that $\ccert(A) \subsetneq B \subseteq \cuniv(A)$, for
    otherwise nothing has to be shown.  We now distinguish two cases:
    \begin{enumerate}[i. ]
    \item $q$ confirms $A \to B$.  Then $\ccert$ is updated to the value of
      \begin{equation*}
        \ccert' = X \mapsto \ccert(\mathcal{L}(\ccert(X)))
      \end{equation*}
      where $\mathcal{L} = \set{A \to B}$ and $X \subseteq M$.  Since $q$ confirms $A \to
      B$ and $\ccert(\cdot) \subseteq \Th(q)(\cdot)$, it follows that $\ccert'(\cdot)
      \subseteq \Th(q)(\cdot)$.

      In the situation before step~\ref{generalae:update-counterexamples}, by
      Lemma~\ref{lemma:counterexamples-are-closed-under-theory} for every element $(C, D)
      \in \con{K}$ it holds that $\Th(q)(C) \cap D = \emptyset$ and hence $\ccert'(C) \cap
      D = \emptyset$.  Moreover, $C' := \ccert'(C)$ is also disjoint to
      \begin{equation*}
        D' := D \cup \set{m \in M \setminus D \mid \ccert'(C \cup \set{m}) \cap D \neq \emptyset}
      \end{equation*}
      and $(C' \to \set{m}) \not\in \Th(q)$ for $m \in D' \setminus D$.  Therefore, after
      step~\ref{generalae:update-counterexamples}, $\Th(q)(C') \subseteq M \setminus D'$
      for every $(C', D') \in \con{K}$.  Then by
      Lemma~\ref{lemma:experts-and-partial-contexts}, $\Th(q)(\cdot) \subseteq
      \con{K}(\cdot)$ and therefore $\ccert'(\cdot) \subseteq \Th(q)(\cdot) \subseteq
      \cuniv(\cdot) \cap \con{K}(\cdot)$ as required.
    \item $q$ gives $(X, Y)$ as a counterexample for $A \to B$.  Then in this iteration
      the value of $\ccert$ is not changed.  The counterexample that is effectively added
      to $\con{K}$ is then
      \begin{equation*}
        (X', Y') = (\ccert(X), Y \cup \set{m \in M \setminus Y \mid \ccert(X \cup \set{m}) \cap Y
          \neq \emptyset}).
      \end{equation*}
      Since $\Th(q)(X') \subseteq M \setminus Y'$, from
      Lemma~\ref{lemma:experts-and-partial-contexts} and the induction hypothesis it
      follows that $\Th(q)(\cdot) \subseteq \con{K}(\cdot)$.  Together with $\Th(q)(\cdot)
      \subseteq \cuniv(\cdot)$ we obtain $\ccert(\cdot) \subseteq \Th(q)(\cdot) \subseteq
      \cuniv(\cdot) \cap \con{K}(\cdot)$ as required.
    \end{enumerate}
  \end{Proof}
\end{Lemma}

We shall at first investigate the already mentioned property that questions asked to the
expert are somehow non-redundant.  We state this kind of non-redundancy as the fact that
the answer to a proposed implication is not predetermined by the current knowledge or by
the answers given so far.

\begin{Theorem}
  \label{thm:general-ae-does-not-ask-redundant-questions}
  Let $M$ be a set, $\ccert, \cuniv$ closure operators on $M$ and $q$ a partial domain
  expert on $M$ such that $\ccert(\cdot) \subseteq \Th(q)(\cdot) \subseteq \cuniv(\cdot)$.
  Suppose that we are in the $n+1$ iteration of
  Algorithm~\ref{algorithm:general-attribute-exploration} and suppose that the implication
  $A \to B$ is asked to the expert $q$.

  Then for each $m \in B$ there exist two partial domain experts $q_1, q_2$ which return
  the same values as $q$ in all iterations $i \in \set{1,\ldots,n}$ and satisfy
  $\ccert(\cdot) \subseteq \Th(q_1)(\cdot), \Th(q_2)(\cdot) \subseteq \cuniv(\cdot)$, such
  that $q_1$ rejects $A \to \set{m}$ and $q_2$ confirms $A \to \set{m}$.
  \begin{Proof}
    Let $\ccert^i, \cuniv^i, \con{K}^i$ be the values of the corresponding closure
    operators and the current working context in iteration $i \in \set{1,\ldots,n}$,
    respectively.  Furthermore, let $A_i \to B_i$ be the implication asked in iteration
    $i$.  Finally, let $\top$ be a symbol not equal to any subset of $M$.

    We then define $q_1$ as follows:
    \begin{equation*}
      q_1(A \to B) =
      \begin{cases}
        q(A \to B) &\text{if } (A \to B) = (A_i \to B_i) \text{ for some } i,\\
        \top       &\text{if } B \subseteq \ccert^n(A),\\
        (\ccert(A), M \setminus \ccert(A)) &\text{otherwise},
      \end{cases}
    \end{equation*}
    for all $(A \to B) \in \Imp(M)$.  Then $q_1$ is a partial domain expert on $M$ and
    $\Th(q_1) = \Th(\ccert^n)$.  Since $\ccert^n(\cdot) \subseteq \cuniv(\cdot)$ by
    Lemma~\ref{lemma:ccert-subseteq-conK} and $m \not\in \ccert^n(A)$, $q_1$ rejects $A
    \to \set{m}$.

    To construct $q_2$ we consider the formal context $\con{K}$ with object set
    $\con{K}^n$, attribute set $M$ and incidence relation $\rel{I}_\con{K}$ given by
    \begin{align*}
      (C, D) \rel{I}_{\con{K}} x &\iff 
      \begin{cases}
        x \in \ccert^n(C \cup \set{m}) &\text{if } m \not\in D\\
        x \in C & \text{otherwise}
      \end{cases}
    \end{align*}
    for all $(C, D) \in \con{K}^n$ and $x \in M$.  By
    step~\ref{generalae:update-counterexamples} in
    Algorithm~\ref{algorithm:general-attribute-exploration}, all object intents of
    $\con{K}$ are closed under $\ccert^n$, therefore $\Th(\ccert^n) \subseteq
    \Th(\con{K})$.  Together with $\ccert(\cdot) \subseteq \ccert^n(\cdot)$ follows
    $\ccert(\cdot) \subseteq \Th(\con{K})(\cdot)$.

    We now define $q_2$ by
    \begin{equation*}
      q_2(A \to B) =
      \begin{cases}
        q(A \to B) &\text{if } (A \to B) = (A_i \to B_i) \text{ for some } i,\\
        \top       &\text{if } B \subseteq A'' \cap \cuniv^n(A), \\
        (X, M \setminus X)
                   &\text{with } X = A'' \cap \cuniv^n(A) \text{ otherwise}
      \end{cases}
    \end{equation*}
    for all $(A \to B) \in \Imp(M)$.  Then $q_2$ is a partial domain expert with $\Th(q_2)
    = \Th(\con{K}) \cap \Th(\cuniv^n)$.  For this we observe that for $(C, D) \in
    \con{K}^n$, if $m \not\in D$, then $\ccert^n(C \cup \set{m}) \cap D = \emptyset$ by
    step~\ref{generalae:update-counterexamples}.  Therefore, the counterexamples given for
    some implication $A_i \to B_i$ from $q$ can also be given by $q_2$.

    Since $\ccert(\cdot) \subseteq \Th(\con{K})(\cdot)$ and $\ccert(\cdot) \subseteq
    \cuniv^n(\cdot)$, it follows that $\ccert(\cdot) \subseteq \Th(q_2)(\cdot) \subseteq
    \cuniv(\cdot)$.

    Furthermore, $m \in \cuniv^n(A)$ and since $\con{K}^n$ does not reject $A \to B$, it
    follows that for each $(C, D) \in \con{K}^n$ with $A \subseteq C$ that $m \not\in D$.
    Hence, $m \in A''$ and therefore $q_2$ confirms $A \to B$ as required.
  \end{Proof}
\end{Theorem}

One of the crucial features of attribute exploration is that it returns a complete
description of the domain of discourse upon termination.  This property does also hold for
our generalized formulation.

\begin{Theorem}
  Let $M$ be a set, $\ccert$, $\cuniv$ closure operators on $M$ and let $q$ be a partial
  domain expert on $M$.  Furthermore, suppose that $\ccert(\cdot) \subseteq \Th(q)(\cdot)
  \subseteq \cuniv(\cdot)$.

  Suppose that Algorithm~\ref{algorithm:general-attribute-exploration} terminates on input
  $\ccert$, $\cuniv$ and $q$ and denote the returned partial context by $\con{K}$ and the
  returned closure operator by $c$.  Let $X \subseteq M$ such that $c(X)$ is finite.
  \begin{enumerate}[i. ]
  \item $\Th(q)(X) = c(X)$.
  \item $c(X) = \cuniv(X) \cap \con{K}(X)$.
  \item Let $\mathcal{K}$ be the set of all implications which have been confirmed by $q$
    during the run of the algorithm.  Define $c'(X)$ to be the smallest set that contains
    $X$ and is closed under both $\ccert$ and $\mathcal{K}(\cdot)$.  Then $c'(X) = c(X)$.
  \item Let $\bar{\con{K}} = (\con{K}, M, I)$ where
    \begin{equation*}
      (C, D) \rel{I} m \iff m \in C.
    \end{equation*}
    Then
    \begin{equation*}
      c(X) = \cuniv(X) \cap X'',
    \end{equation*}
    where $(\cdot)''$ denotes the double derivation operator in $\bar{\con{K}}$.
  \end{enumerate}

  \begin{Proof}
    By Lemma~\ref{lemma:ccert-subseteq-conK}, $\ccert'(\cdot) \subseteq \Th(q)(\cdot)
    \subseteq \cuniv'(\cdot)$ holds at the end of every iteration in the run of the
    algorithm, where $\ccert'$ and $\cuniv'$ denote the current values of the
    corresponding closure operators.  Since the algorithm terminates, $\ccert'(Y) =
    \cuniv'(Y)$ holds in the last iteration for all $Y \subseteq M$ if $\ccert'(Y)$ is
    finite.  Since $c = \ccert'$ and $\ccert'(X) \subseteq \Th(q)(X) \subseteq
    \cuniv'(X)$, the first assertion follows.

    By induction on the number of iterations of the algorithm, one can see that at the end
    of every iteration of the algorithm it holds that $\cuniv'(X) = \cuniv(X) \cap
    \con{K}(X)$, where $\cuniv'$ is the current value of the upper closure operator,
    $\cuniv$ is the original value of the upper closure operator and $\con{K}$ is the
    current working context.  Since the algorithm terminates, $\cuniv'(X) = c(X)$ holds in
    the last iteration and the second claim follows.

    Suppose that the algorithm is in a certain iteration and suppose that $\mathcal{K}'$
    is the set of confirmed implications up to now.  By induction we see that if $\ccert'$
    is the current value of the lower closure operator, then $\ccert'(X)$ is the smallest
    set containing $X$ that is closed both under $\ccert$ and $\mathcal{K}'(\cdot)$.  As
    $c$ is the last value of the lower closure operator during the run of the algorithm,
    $c(X) = c'(X)$, which shows the third claim.

    For the last claim we observe the following relations:
    \begin{align*}
      X''
      &= \bigcap_{(C, D) \in \con{K}, X \subseteq C} C\\
      &\subseteq \bigcap_{(C, D) \in \con{K}, X \subseteq C} M\setminus D\\
      &= \con{K}(X).
    \end{align*}
    By step~\ref{generalae:update-counterexamples} of the algorithm, $C$ is closed under
    $c$ for every $(C, D) \in \con{K}$.  Therefore, $c(X) \subseteq X''$.  Together this
    yields
    \begin{equation*}
      \cuniv(X) \cap c(X) \subseteq \cuniv(X) \cap X'' \subseteq \cuniv(X) \cap \con{K}(X)
    \end{equation*}
    and since $c(X) \subseteq \cuniv(X)$ and $c(X) = \cuniv(X) \cap \con{K}(X)$, the last
    claim follows.
  \end{Proof}
\end{Theorem}

Termination of the generalized attribute exploration algorithm is not guaranteed in
general (\ie when $M$ is infinite and $\ccert$ and $\cuniv$ are arbitrary).  Hence,
termination normally has to be shown for the concrete application at hand.  We can,
however, give some sufficient condition which may still be helpful.

\begin{Theorem}
  The general attribute exploration algorithm with input $\ccert$, $\cuniv$ and a partial
  domain expert $q$ terminates if there are only finitely many closure operators $c$ on
  $M$ such that $\ccert(\cdot) \subsetneq c(\cdot) \subsetneq \cuniv(\cdot)$.
  \begin{Proof}
    The claim follows easily if we can show that in every iteration of attribute
    exploration either the value of $\ccert$ is updated to a new value $\ccert'$ such that
    $\ccert \subsetneq \ccert' \subseteq \cuniv$ or, likewise, if the value for $\cuniv$
    is updated to a new value $\cuniv'$ such that $\ccert \subseteq \cuniv' \subsetneq
    \cuniv$.

    Let $A$ be such that $\ccert(A) \neq \cuniv(A)$ and let $B \subseteq M$ be finite such
    that $\ccert(A) \subsetneq B \subseteq \cuniv(A)$.  If $q$ confirms $A \to B$, then
    $\ccert$ is updated to the value
    \begin{equation*}
      \ccert'(X) = \ccert(\mathcal{L}(\ccert(X))),
    \end{equation*}
    where $\mathcal{L} = \set{A \to B}$ and $X \subseteq M$.  Clearly, $\ccert(\cdot) \subsetneq
    \ccert'(\cdot)$ and by Lemma~\ref{lemma:ccert-subseteq-conK}, $\ccert'(\cdot) \subseteq \cuniv(\cdot)$.

    If $q$ yields a counterexample $(C, D)$ for $A \to B$, then the new value $\cuniv'$
    for $\cuniv$ is computed by
    \begin{equation*}
      \cuniv'(X) = \cuniv(X) \cap \con{K}(X)
    \end{equation*}
    for $X \subseteq M$.  It follows that $\cuniv'(\cdot) \subseteq \cuniv(\cdot)$ and $\cuniv'(A)
    \subseteq \cuniv(A) \setminus D \subsetneq \cuniv(A)$, since $C \subseteq A$, $B
    \subseteq \cuniv(A)$ and $B \cap D \neq \emptyset$.  By
    Lemma~\ref{lemma:ccert-subseteq-conK} it follows that $\ccert(X) \subseteq \con{K}(X)$
    for all $X \subseteq M$.  Hence $\ccert(\cdot) \subseteq \cuniv'(\cdot) \subsetneq \cuniv(\cdot)$ as
    required.
  \end{Proof}
\end{Theorem}

Of course, if after finitely many iterations the situation of the theorem is reached, the
generalized attribute exploration will terminate as well.

\section{Computing Undecided Implications}
\label{sec:computing-undecided-implications}

We have seen that a lot of the useful properties of attribute exploration remain true in
our generalized form of Algorithm~\ref{algorithm:general-attribute-exploration}.  However,
we have not discussed the property of the classical attribute exploration that the number
of questions which the expert confirms is minimal.  Indeed, we cannot expect that from our
generalization, as we have not opposed any restriction on the order in which implications
are asked.  It is therefore possible to ask an implication $A \to B$, which is confirmed,
just to ask in the next iteration an implication $A \to C$ with $C \supseteq B$, which
might also get confirmed.  It is therefore advisable to always ask implications with
$\subseteq$-maximal conclusions.  However, even in that case it might not be clear whether
the number of confirmed implications asked is really minimal.

We therefore want to discuss in this section whether it is possible to modify our general
attribute exploration such that the number of questions asked such that the expert
confirms is the smallest possible.  For this we shall try to adapt the computation of
undecided implications from the classical case.

Let us recall how implications asked to a domain expert $p$ are computed in the case of
classical attribute exploration, as discussed in
Algorithm~\ref{algorithm:classical-attribute-exploration}.  For this suppose that we are
in a certain iteration of the algorithm, with known implications $\mathcal{K}$, working
context $\con{K}$ and $P$ the last computed premise.  Further suppose that we have fixed a
total order on the set $M$ before the start of the algorithm, which induces a lectic order
$\preceq$ on $\subsets{M}$.  Then, in the classical case, we compute the lectically
smallest set $Q \subseteq M$ after $P$ that is closed under $\mathcal{K}$ and that is not
an intent of $\con{K}$.  The implication $Q \to Q''$ is then asked to $p$.

Computing the lectically next set after a set $P$ can be done using the Next-Closure
algorithm~\cite{DBLP:conf/icfca/Ganter10}.  However, for theoretical considerations we can
neglect lectic orderings, as we shall see in a moment.

Let $M$ be a finite set.  To guarantee that the number of confirmed implications is as
small as possible, we change step~\ref{generalae:undecided-implication} to:
\begin{itemize}
\item[\ref*{generalae:undecided-implication}'. ] Let $A \subseteq M$ be such that $A =
  \ccert(A) \subsetneq \cuniv(A)$ and $A$ is $\subseteq$-minimal with respect to this
  property.  Consider the implication $A \to \cuniv(A)$.
\end{itemize}
This is a generalization of the corresponding step in the classical case.  If $P$ is the
premise of the last implication asked, then the lectically next set $Q$ after $P$ is a
$\subseteq$-minimal set with $Q = \mathcal{K}(Q) \subsetneq Q''$, and the implication $Q
\to Q''$ is asked next.

Before we give the formal statement of the fact that this indeed yields an algorithm that
always asks a minimal number of confirmed implications, we shall give the following
definition.

\begin{Definition}
  Let $c_1, c_2$ be two closure operators on a finite set $M$ and let $P \subseteq M$.
  Then $P$ is said to be \emph{$c_1$-pseudoclosed under $c_2$} if and only if
  \begin{enumerate}[i. ]
  \item $c_1(P) = P$,
  \item $c_2(P) \neq P$,
  \item for all $Q \subsetneq P$ being $c_1$-pseudoclosed under $c_2$ it follows that
    $c_2(Q) \subseteq P$.
  \end{enumerate}
\end{Definition}

\begin{Theorem}
  Consider Algorithm~\ref{algorithm:general-attribute-exploration} with
  step~\ref{generalae:undecided-implication} replaced by
  step~\ref{generalae:undecided-implication}'.

  Let $M$ be a finite set, $q$ a partial domain expert on $M$, $\ccert$, $\cuniv$ closure
  operators on $M$ such that $\ccert(\cdot) \subseteq \Th(q)(\cdot) \subseteq
  \cuniv(\cdot)$.  Let $\mathcal{K}$ be the set of confirmed implications during the run
  of the algorithm with input $\ccert$, $\cuniv$ and $q$, and let $c$ be the returned
  closure operator.

  Then the premises of the implications in $\mathcal{K}$ are exactly the
  $\ccert$-pseudoclosed sets of $c$.
  \begin{Proof}
    We show that a set $A \subseteq M$ is a $\ccert$-pseudoclosed set of $c$ if and only
    if the implication $A \to c(A)$ is asked to and confirmed by $q$.  We shall do so
    using well-founded induction, which is possible since $M$ is finite.

    Let $A$ be a premise of a confirmed implication $A \to B$.  It follows that $B =
    \cuniv'(A)$ for the corresponding value of $\cuniv'$ in the iteration in which $A \to
    B$ is asked to $q$.  Then $A$ is closed under $\ccert$ and under all currently known
    implications, \ie under
    \begin{equation*}
      \set{X \to Y \mid (X \to Y) \in \mathcal{K}, X \subseteq A}.
    \end{equation*}

    Suppose that their exists an implication $(X \to Y) \in \mathcal{K}$ such that $X
    \subseteq B$.  Then $Y \subseteq \cuniv'(B) = \cuniv'(A) = B$.  Therefore, $B$ is
    closed under $\mathcal{K}$ and hence $B = c(A)$.

    We shall show next that $A$ is a $\ccert$-pseudoclosed set of $c$.  We already know
    that $A$ is closed under $\ccert$.  Furthermore, since $A \to B$ is asked to $q$, $B
    \neq A$ and therefore $A \neq c(A)$.

    Let $R \subsetneq A$ be a $\ccert$-pseudoclosed set of $c$.  By the induction
    hypothesis, $R \to c(R)$ is asked to and confirmed by $q$.  Since $A$ is closed under
    all those implications, it follows that $c(R) \subseteq A$ as required.

    Conversely, let $A$ be a $\ccert$-pseudoclosed set of $c$.  By the induction
    hypothesis, for all $\ccert$-pseudoclosed sets $R \subsetneq A$ the implication $R \to
    c(R)$ is asked to and confirmed by $q$.  Since $c(R) \subseteq A$ and $\ccert(A) = A$
    it follows that $A$ is $\subseteq$-minimal with respect to being closed under $\ccert$
    and all confirmed implications $X \to Y$ with $X \subseteq A$.  Therefore, $A \to
    \cuniv'(A)$ will be asked in a certain iteration, with the corresponding value of
    $\cuniv'$.  Since $c(A) \subseteq \cuniv'(A)$ and $A \neq c(A)$, after a finite number
    of counterexamples $A \to c(A)$ will be asked to and confirmed by $q$.
  \end{Proof}
\end{Theorem}

Recall the fact that the set
\begin{equation*}
  \mathcal{K} := \set{P \to c(P) \mid P \text{ is $\ccert$-pseudoclosed set of $c$}}
\end{equation*}
has minimal cardinality such that every set $A \subseteq M$ is closed under $c$ if and
only if $A$ is closed under $\ccert$ and $\mathcal{K}$.  This has been proven
in~\cite{Diss-Felix} for the case of $\ccert = \mathcal{K}(\cdot)$ for a set $\mathcal{K}
\subseteq \Imp(M)$ and $c = (\cdot)''$ for some given formal context $\con{K}$ with
$\con{K} \models \mathcal{K}$.  However, the proof given there also holds in our general
setting.

Summing up, we obtain our desired result.

\begin{Corollary}
  The number of confirmed implications during the run of the general attribute exploration
  algorithm is as small as possible.
\end{Corollary}

\section{Conclusions}

Starting from a classical formulation of attribute exploration using domain experts, we
have presented a more general formulation of attribute exploration that is able to work
with abstractly given closure operators and can handle partially given counterexamples.
We have also seen that most of the properties of classical attribute exploration remain in
general or, as in the case of minimality of confirmed implications, under certain
restrictions.


\bibliography{fca,dl}

\begin{thebibliography}{10}

\bibitem{BaaderDistel08}
Franz Baader and Felix Distel.
\newblock A finite basis for the set of {EL}-implications holding in a finite
  model.
\newblock In Raoul Medina and Sergei Obiedkov, editors, {\em Proceedings of the
  6th International Conference on Formal Concept Analysis, (ICFCA 2008)},
  volume 4933 of {\em Lecture Notes in Artificial Intelligence}, pages 46--61.
  Springer Verlag, 2008.

\bibitem{BaDi09}
Franz Baader and Felix Distel.
\newblock Exploring finite models in the description logic
  $\mathcal{EL}_\mathrm{gfp}$.
\newblock In S\'ebastien Ferr\'e and Sebastian Rudolph, editors, {\em
  Proceedings of the 7th International Conference on {F}ormal {C}oncept
  {A}nalysis, {(ICFCA 2009)}}, volume 5548 of {\em Lecture Notes in Artificial
  Intelligence}, pages 146--161. Springer Verlag, 2009.

\bibitem{BGSS07}
Franz Baader, Bernhard Ganter, Ulrike Sattler, and Baris Sertkaya.
\newblock Completing description logic knowledge bases using formal concept
  analysis.
\newblock In {\em Proceedings of the Twentieth International Joint Conference
  on Artificial Intelligence {(IJCAI-07)}}, pages 230--235. AAAI Press, 2007.

\bibitem{DBLP:conf/icfca/Distel10}
Felix Distel.
\newblock {H}ardness of {E}numerating {P}seudo-intents in the {L}ectic {O}rder.
\newblock In Kwuida and Sertkaya \cite{DBLP:conf/icfca/2010}, pages 124--137.

\bibitem{Diss-Felix}
Felix Distel.
\newblock {\em {Learning Description Logic Knowledge Bases from Data Using
  Methods from Formal Concept Analysis}}.
\newblock PhD thesis, {TU Dresden}, 2011.

\bibitem{DBLP:journals/tcs/Ganter99}
Bernhard Ganter.
\newblock Attribute exploration with background knowledge.
\newblock {\em Theor. Comput. Sci.}, 217(2):215--233, 1999.

\bibitem{DBLP:conf/icfca/Ganter10}
Bernhard Ganter.
\newblock Two basic algorithms in concept analysis.
\newblock In Kwuida and Sertkaya \cite{DBLP:conf/icfca/2010}, pages 312--340.

\bibitem{GORS-book}
Bernhard Ganter, Sergei Obiedkov, Sebastian Rudolph, and Gerd Stumme.
\newblock {Conceptual Exploration}.
\newblock to appear.

\bibitem{fca-book}
Bernhard Ganter and Rudolph Wille.
\newblock {\em Formal Concept Analysis: Mathematical Foundations}.
\newblock Springer, Berlin-Heidelberg, 1999.

\bibitem{DBLP:conf/icfca/2010}
L{\'e}onard Kwuida and Baris Sertkaya, editors.
\newblock {\em Formal Concept Analysis, 8th International Conference, ICFCA
  2010, Agadir, Morocco, March 15-18, 2010. Proceedings}, volume 5986 of {\em
  Lecture Notes in Computer Science}. Springer, 2010.

\bibitem{stumme96attribute}
Gerd Stumme.
\newblock Attribute exploration with background implications and exceptions.
\newblock In H.-H. Bock and W.~Polasek, editors, {\em Data Analysis and
  Information Systems. Statistical and Conceptual approaches. Proc. GfKl'95.
  Studies in Classification, Data Analysis, and Knowledge Organization 7},
  pages 457--469, Heidelberg, 1996. Springer.

\end{thebibliography}
\bibliographystyle{plain}

\end{document}